\documentclass[12pt]{article}

\usepackage[pdftex,usenames,dvipsnames]{color}
\usepackage{graphics}
\usepackage{a4wide}
\usepackage{amsfonts}
\usepackage{amsmath}	
\usepackage{amssymb}

\newcommand{\nit}{\noindent}

\newcommand{\np}{\newpage}
\newcommand{\dsp}{\displaystyle}
\newcommand{\vs}[1]{\vspace{#1 ex}}
\newcommand{\hs}[1]{\hspace{#1 em}}
\newcommand{\bfr}{\begin{flushright}}
\newcommand{\efr}{\end{flushright}}
\newcommand{\bc}{\begin{center}}
\newcommand{\ec}{\end{center}}
\newcommand{\ben}{\begin{enumerate}}
\newcommand{\een}{\end{enumerate}}

\newcommand{\be}{\begin{equation}}
\newcommand{\ee}{\end{equation}}
\newcommand{\ba}{\begin{array}}
\newcommand{\ea}{\end{array}}
\newcommand{\ct}{\cite}
\newcommand{\bit}{\bibitem}

\newcommand{\ag}{\alpha}

\newcommand{\gam}{\gamma}
\newcommand{\del}{\delta}

\newcommand{\ve}{\varepsilon}

\newcommand{\vf}{\varphi}

\newcommand{\Del}{\Delta}

\newcommand{\lh}{\left(}
\newcommand{\rh}{\right)}
\newcommand{\ld}{\left.}
\newcommand{\rd}{\right.}

\newcommand{\ctg}{\mbox{\,cotan}}

\newcommand{\arccotanh}{\mbox{\,arccotanh}}

\newcommand{\bea}{\begin{eqnarray}}
\newcommand{\beas}{\begin{eqnarray*}}
\newcommand{\eea}{\end{eqnarray}}
\newcommand{\eeas}{\end{eqnarray*}}

\newcommand{\schwa}{1-\frac{2M}{r}}
\newcommand{\bes}{\begin{equation*}}
\newcommand{\ees}{\end{equation*}}

\begin{document}

\pagestyle{empty}

\bfr
NIKHEF/2014-020
\efr

\bc
{\Large {\bf Ballistic orbits in Schwarzschild space-time  }}\\
\vs{3}

{\Large {\bf and gravitational waves from EMR binary mergers}}\\
\vs{7}

{\large G.\ d'Ambrosi$^a$  and J.W.\ van Holten$^b$ }\\
\vs{3}

{\large NIKHEF} \\

{\large Amsterdam NL} \\
\vs{3}

July 2, 2014

\ec
\vs{10}

\nit
{\footnotesize 
{\bf Abstract} \\
We describe a special class of ballistic geodesics in Schwarzschild space-time, extending to the horizon in the infinite past
and future of observer time, which are characterized by the property that they are in 1-1 correspondence, and completely 
degenerate in energy and angular momentum, with stable circular orbits. We derive analytic expressions for the source terms 
in the Regge-Wheeler and Zerilli-Moncrief equations for a point-particle moving on such a ballistic orbit, and compute the 
gravitational waves emitted during the infall in an Extreme Mass Ratio black-hole binary coalescence. In this way a geodesic 
approximation to the plunge phase of compact binaries is obtained.}

\vfill

\footnoterule
{\footnotesize  \hs{-2.1} $^a$ e-mail: gdambros@nikhef.nl \\ $^b$ e-mail: v.holten@nikhef.nl}

\np
 ~\hfill
 
\np
\pagestyle{plain}
\pagenumbering{arabic}

\section{Introduction}
The physics of black holes has become a subject of intense interest, pursued in theoretical and observational 
research by physicists and astronomers. One of the most promising roads to explore large black holes is by the 
detection of gravitational radiation from their interaction with compact objects, including other black holes. 
 
It is widely expected that in the next decade the detection of gravitational waves of astrophysical origin will 
become a reality. Experimental evidence is to come from the ground-based interferometers LIGO and VIRGO 
\cite{Somiya,Accadia} as well as from the future missions eLISA and DECIGO \cite{Seoane}. Binary systems
including a black hole are among the most prominent potential sources for detection. 

The class of Extreme Mass-Ratio binaries (EMRs), in which a black hole of large mass $M$ is accompanied 
by a compact object of much smaller mass $\mu$, such that $\mu/M \ll 1$, is of particular interest because
the compact object acts in many respects as a point-like probe of the black-hole geometry. From the theoretical 
point of view, compact binaries like these can be studied via perturbative methods, with the static or stationary
black-hole solutions of General Relativity as a first-order approximation \cite{2bodyreview,Schultz}. 
 
Analytical results for the orbits of test masses in Schwarzschild space-time are found in standard text 
books \ct{chandra, MTW, hartle}. Complete geodesics --meaning the full space-time position as a function 
of (proper) time-- are less easy to compute, and usually given only in the form of implicit expressions. 
For practical applications several perturbative schemes have been developed to construct satisfactory 
approximations. The post-newtonian expansion, computing relativistic corrections to Kepler orbits, has 
been developed to high order in the parameters\footnote{In the remainder of this paper we use natural units 
$c = G = 1$.} $v/c$ and $GM/rc^2$; see refs.\ \ct{PNoverview,Maggiore} and work cited there. In refs.\ 
\ct{OriginalEpicycle,GKJWvH, GKJWvH2,koekoek} a different, fully relativistic scheme has been developed 
based on covariant deformation of known (circular) orbits. This last method, developed to second order in 
the deformation parameter, was also shown to give very good results for the gravitational wave signals from 
quasi-periodic bound orbits with moderate eccentricities.  

In the present paper we extend the relativistic calculations to include a particular class of unstable orbits, 
describing test masses falling towards the horizon in the regime where the motion is no longer quasi-periodic. 
A well-known extreme case of infall is the straight plunge along a radial orbit, which was dealt with extensively 
in refs.\ \ct{MartelPoisson, Martel}. Here we focus on the opposite extreme, infall from a periodic orbit, in 
particular circular orbits close to the innermost stable circular orbit (ISCO).

Such unstable orbits are interesting on their own and shed light on the dynamics of EMRs, so they can also be used 
for the evaluation of gravitational radiation. It is known \ct{Buonanno} that binary coalescences go through three 
different stages, which are commonly referred to as the inspiral, plunge and merger-ringdown. During the 
\emph{inspiral} of an EMR binary, the smaller companion $\mu$ follows bound quasi-periodic orbits, such as those 
studied in \ct{GKJWvH, GKJWvH2,koekoek}. Loss of energy through GWs generally translates in a loss of eccentricity 
(\emph{circularisation})  in this phase, although the eccentricity may increase again just before plunging \cite{Cutler}, 
depending on the parameters of the orbit and its evolution\footnote{However, the relative increase of the eccentricity
is modest: $e \ll 1$, while during previous stages of the inspiral $e$ can drop by orders of magnitude.}. However, the 
region of the ISCO is where the \emph{plunge} sets in and the unstable orbits in some range of parameters can be 
used in the description of this phase and the following. In particular they show that an EMR plunge can follow an 
almost-circular path at the beginning, up to radial values $r \sim 4.3 M$, when $\dot{r}\sim r\dot{\varphi}$ and the
radial velocity becomes comparable to the transverse velocity, after which the radial motion becomes dominant.
Moreover our particular unstable orbits may be the starting point for a geodesic deviation expansion 
\cite{GKJWvH,GKJWvH2} which can give an even more realistic description of the final plunge of an EMR system. 
 
This paper is organized as follows. In sect.\ 2 we review some elements of geodesic motion in Schwarzschild 
space-time, and show the existence of unstable infalling geodesics which are completely degenerate in energy 
and angular momentum per unit of mass with circular orbits. In particular, just like for the circular orbits (but 
approaching from the other side) the limit of stability of these geodesics is the ISCO. In sect.\ 3 we give explicit 
analytic relations between the space-time co-ordinates $(t, r, \vf)$ of an infalling test mass, including an expression 
for the proper time $\tau$. In sect.\ 4 we use these results as input to derive expressions for the source terms in 
the Regge-Wheeler and Zerilli-Moncrief equations \ct{RW,zerilli,moncrief} for gravitational waves of a point mass 
in a Schwarzschild background. The wave equations are solved numerically using the Lousto-Price algorithm 
\ct{lousto-price}. In the final section we discuss the results and compare them with similar results in the literature 
for infall from the ISCO. 

\section{Motion in Schwarzschild space-time} 

Time-like geodesics in the equatorial plane of Schwarzschild space-time are characterized by two constants of 
motion, the proper energy and angular momentum per unit of mass of a test particle:
\be
\ve = \lh 1 - \frac{2M}{r} \rh \frac{dt}{d\tau}, \hs{2} \ell = r^2\, \frac{d\vf}{d\tau}.
\label{2.1}
\ee
In terms of these the four-velocity constraint $u_{\mu} u^{\mu} = -1$ is equivalent to
\be
\ve^2 = \lh \frac{dr}{d\tau} \rh^2 + V_{\ell}(r),
\label{2.2}
\ee
with the effective potential
\be
V_{\ell} = \lh 1 - \frac{2M}{r} \rh \lh 1 + \frac{\ell^2}{r^2} \rh.
\label{2.3}
\ee
It follows, that for $\ell^2 > 12 M^2$ there exist {\em circular} orbits with constant radial co-ordinate $r$. In fact, 
there are two types of circular orbits, a stable one at $r = R_+$ in the minimum of $V_{\ell}$, and an unstable 
one at $r = R_-$ in its maximum, with
\be
R_{\pm} = \frac{\ell^2}{2M} \lh 1 \pm \sqrt{1 - \frac{12M^2}{\ell^2}} \rh.
\label{2.4}
\ee
For $R = 6M$ and $\ell^2 = 12 M^2$ the effective potential $V_{\ell}$ only has a point of inflection, and these two 
solutions coincide; this is the well-known innermost stable circular orbit (ISCO). For $\ell^2 < 12 M^2$ the potential 
$V_{\ell}$ has no stable points for any $\ell$, and no circular orbits exist.  
The expression for $V_{\ell}$ for different circular orbits is conveniently parametrized by
\be
\xi = \sqrt{1 - \frac{12M^2}{\ell^2}} \hs{1} \Rightarrow \hs{1} 
\ve_{\pm}^2 = V_{\ell}[R_{\pm}] = \frac{2}{9} \frac{(2 \pm \xi)^2}{1 \pm \xi}.
\label{2.5}
\ee 
Note that for orbits near the ISCO (small $\xi$), both the energy and angular momentum of circular orbits vary only 
as $\xi^2$. As a result, for orbits near the ISCO the effective potential is very flat between $R_-$ and $R_+$; 
in fact, in the domain $\xi < 0.3$ the difference between maximum and minimum of $V_{\ell}$ is less than 1 percent. 
\vs{1}

\bc
\scalebox{0.25}{\includegraphics{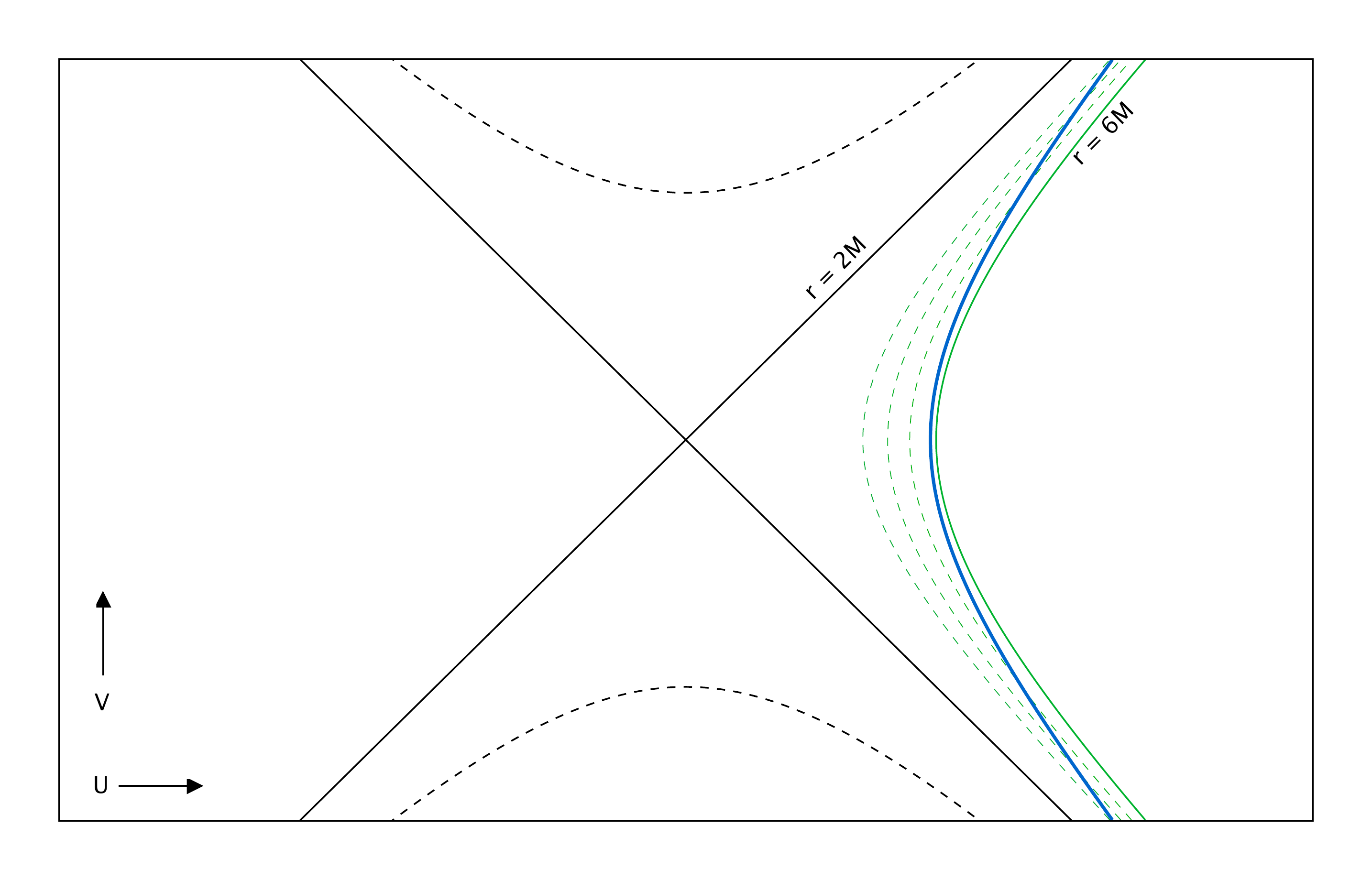}}
\vs{1}

{\footnotesize{Fig.\ 1: Kruskal diagram showing a ballistic orbit crossing lines of constant $r < 6M$ (dotted curves).
                              The ISCO $(r = 6M)$ is the ultimate limit of ballistic orbits.}}
\ec

\nit
In this paper we consider a second class of unstable orbits. These orbits lie completely within the ISCO, 
starting from and falling back into the horizon; therefore we refer to them as {\em ballistic} orbits. An
example is shown in fig.\ 1. We are especially interested in the class of orbits in 1-1-correspondence 
with circular geodesics, with the simple analytic representation 
\be
r = \frac{R}{1 + e \ctg^2 (A \vf/2)}.
\label{2.7}
\ee
Here $(R,e,A)$ are constants fixed by the values of $\ve$ and $\ell$. More precisely, just like for 
circular orbits, there is only one independent parameter fixing a ballistic orbit, which we can choose 
to be the extremal point (apastron) $r  = R$ when $A \vf = \pi$. Then the other constants defining the 
ballistic orbit are
\be
\ba{l}
\dsp{ A^2 = \frac{1}{2} \lh \frac{6M}{R} - 1 \rh, \hs{2} e = \frac{3}{2} \lh 1 - \frac{R}{6M} \rh, }\\
 \\
\dsp{ \ve^2 = \frac{\lh 1 + \frac{2M}{R} \rh^2}{1 + \frac{6M}{R}}, \hs{2} 
  \ell^2 = \frac{16M^2}{\lh 1 - \frac{2M}{R} \rh \lh 1 + \frac{6M}{R} \rh}. }
\ea
\label{2.8}
\ee
It follows that these orbits only exist for $2M < R \leq 6M$, and the limiting case $R = 6M$ is just the
ISCO itself, with $e = A = 0$. For these orbits the value of the parameter $\xi$, defined in (\ref{2.5}), is 
\be
\xi = A^2 = \frac{1}{2} \lh \frac{6M}{R} - 1 \rh,
\label{2.9}
\ee
from which it follows that
\be
\ve^2 = \frac{2}{9} \frac{(2 + \xi)^2}{1+ \xi}.
\label{2.10}
\ee
Thus it is established that these ballistic orbits inside the ISCO are degenerate in energy and angular 
momentum with the stable circular orbits outside the ISCO; see fig.\ 2. 

\bc
\scalebox{0.7}{\includegraphics{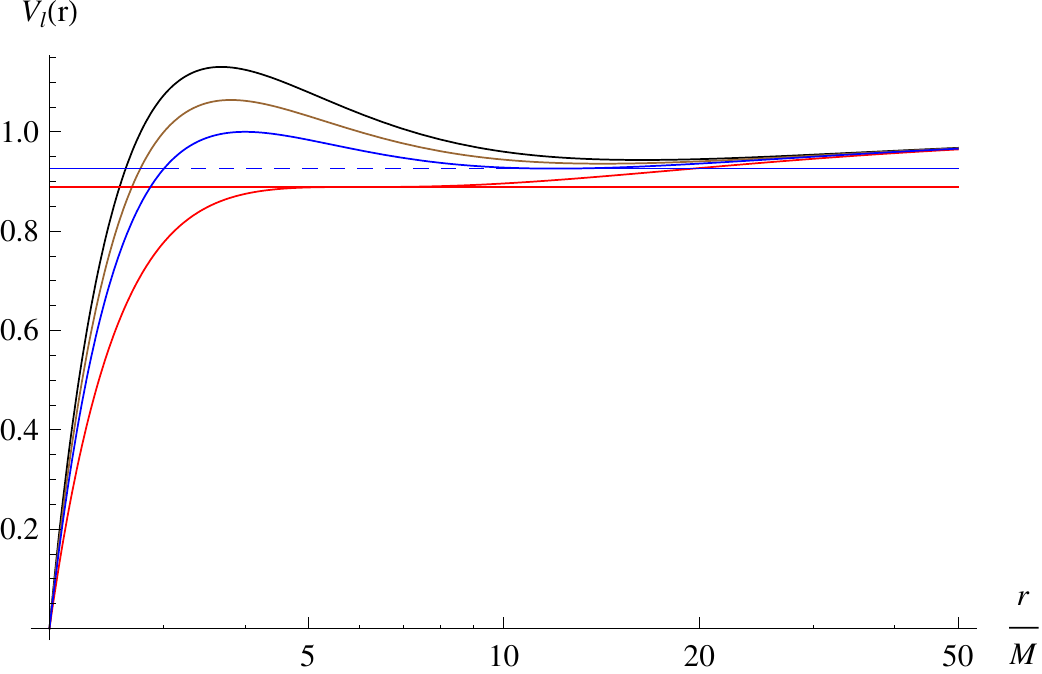}}
\vs{1}

{\footnotesize Fig.\ 2: $V_{\ell}(r)$ for several $\ell^2 \geq 12M^2$, where the equality sign determines the ISCO 
(lowest curve). The horizontal lines indicate the energy levels of the stable circular orbits (dashed) and corresponding 
ballistic orbits (continuous).}
\ec

\nit
In summary, we have analytic representations for three kinds of special geodesics: stable and unstable circular orbits 
with respectively
\be
R_+ = \frac{6M}{1- \xi}, \hs{2} R_- = \frac{6M}{1 + \xi}, 
\label{2.11}
\ee
and special ballistic orbits with apastron
\be
R = \frac{6M}{1 + 2\xi}.
\label{2.12}
\ee
These ballistic orbits are always degenerate in energy and angular momentum with the stable circular orbit
for the same $\xi$, whilst the unstable circular orbit at this $\xi$ has a slightly higher energy; however, all three 
become degenerate at the ISCO where $\xi = 0$. Other ballistic orbits exist; an analytic representation for these 
can be obtained from the ones given in eqs.\ (\ref{2.7}) and (\ref{2.8}) by the method of geodesic deviations 
\ct{bhk,OriginalEpicycle}.

\section{Infall on a ballistic orbit}

As equation (\ref{2.7}) is invariant under $\vf \rightarrow - \vf$, the ballistic orbit  is symmetric about the apastron
$r = R$. The second half of the orbit describes the infall of a test mass from the apastron to the horizon. Now the 
degeneracy in energy and angular momentum of the ballistic orbits and circular orbits implies that  in both cases 
these quantities are greater than those of the ISCO. Thus it follows, that when a test particle in stable motion on 
the ISCO is slightly boosted with the correct angular momentum, it can either move up to a larger distance from 
the horizon on a stable circular orbit, or move down to such an infalling ballistic orbit of the same $\ve$ and $\ell$. 
In real systems such a small boost could happen for instance by interaction with a third body passing by.

In this section we describe the infalling orbit in some more detail. First note, that eq.\ (\ref{2.7}) expresses the
radial co-ordinate $r$ as a function of $\vf$. By using the constants of motion (\ref{2.1}) we can also determine 
the explicit functional dependence of the co-ordinate time $t$ and the proper time $\tau$ on the angle $\vf$. 
The expression for proper time is the simplest one; it is obtained by integrating the relation
\be
\frac{d\vf}{d\tau} = \frac{\ell}{r^2}= \frac{\ell}{R^2}\, \lh 1 + e \ctg^2 \frac{A\vf}{2} \rh^2,
\label{n3.0}
\ee
with the result
\be
\ba{lll}
\dsp{ \lh \frac{1 - e}{(3-e)(3 - 2e)} \rh^{3/2} \frac{\tau - \tau_0}{2M} }& = & \dsp{
 \frac{A \vf}{(3-e) \sqrt{e}} - \arctan \lh \frac{1}{\sqrt{e}} \tan \frac{A\vf}{2} \rh }\\
 & & \\
 & & \dsp{ +\, \frac{\sqrt{e} (1-e)}{3-e}\, \frac{\ctg \frac{A\vf}{2}}{1 + e \ctg^2 \frac{A\vf}{2}}. }
\ea
\label{n3.1}
\ee
Here $\tau_0$ is a constant of integration fixing the zero point of proper time. A convenient choice is to take
$\tau = 0$ at $r=R$.  Similarly, we can solve for $t$ as a function of $\vf$ by integrating
\be
\ba{lll}
\dsp{ \frac{dt}{d\vf} }& = & \dsp{ \frac{\ve}{\ell}\, \frac{r^3}{r - 2M} }\\
 & & \\
 & = & \dsp{ \frac{2MR\ve}{\ell} \lh \frac{R}{2M} \frac{1}{\lh 1 + e \ctg^2 \frac{A\vf}{2} \rh^2} 
 + \frac{1}{1 + e \ctg^2 \frac{A\vf}{2}} + \frac{1}{\frac{R}{2M} - 1 - e \ctg^2 \frac{A\vf}{2}} \rh. }
\ea
\label{n3.3}
\ee

\nit
The result is
\be
\ba{lll}
\dsp{ \frac{1}{(2-e)} \sqrt{\frac{e}{2(1-e)}}\, \frac{t - t_0}{4M} }& = & \dsp{
  \frac{(3 - 2e)^2 A \vf}{2(2-e)(1 - e)^2} + \frac{e(3 - 2e)}{2(1 - e)} \frac{\ctg \frac{A\vf}{2}}{1 + e \ctg^2 \frac{A\vf}{2}} }\\
 & & \\
 & & \dsp{ -\, \frac{(11 - 11e + 2e^2) \sqrt{e}}{2(1 - e)^2} \arctan \lh \frac{1}{\sqrt{e}} \tan \frac{A\vf}{2} \rh }\\
 & & \\
 & & \dsp{ -\, \frac{1}{(2-e)} \sqrt{\frac{e}{2(1-e)}} \arccotanh \lh \sqrt{\frac{2(1-e)}{e}} \tan \frac{A\vf}{2} \rh. }
\ea
\label{n3.4}
\ee
Finally, this result can be translated to a result for $t$ as a function of $r$:
\be
\ba{l}
\dsp{ \frac{2}{(2-e)}\sqrt{\frac{e}{2(1-e)}}\frac{t-t_0}{4M} =  
 \frac{\left(11-11 e+2 e^2\right)\sqrt{e}}{(1-e)^2}\, \arctan \sqrt{\frac{r}{(6-4 e) M-r}} }\\
 \\
\dsp{ \hs{3} -\, \frac{2 (3-2 e)^2 }{(1-e)^2 (2-e)}\, \arctan \sqrt{\frac{e r}{(6-4 e) M-r}} 
  - \frac{ \sqrt{e r \lh (6-4 e)M - r \rh}}{2M (1-e)} }\\
  \\
 \dsp{ \hs{3} +\, \frac{1}{2-e} \sqrt{\frac{2e}{1-e}}\, \arccotanh  \sqrt{\frac{2(1-e) r}{(6-4 e) M-r}}.}
\ea
\label{n3.5}
\ee

\nit
A typical orbit close to the ISCO, with $e = 1.5 \times 10^{-3}$ is shown in fig.\ 3. This orbit encircles the
black hole about 20 times before crossing the horizon. Moreover, the radial motion is seen to be small 
compared to the transverse motion until the radial co-ordinate gets in the domain $4M < r  < 5M$. Once the 
cross-over to radially dominated motion is made, only a few turns remain. These observations will be 
made more precise in the following. 

\bc
\scalebox{0.6}{\includegraphics{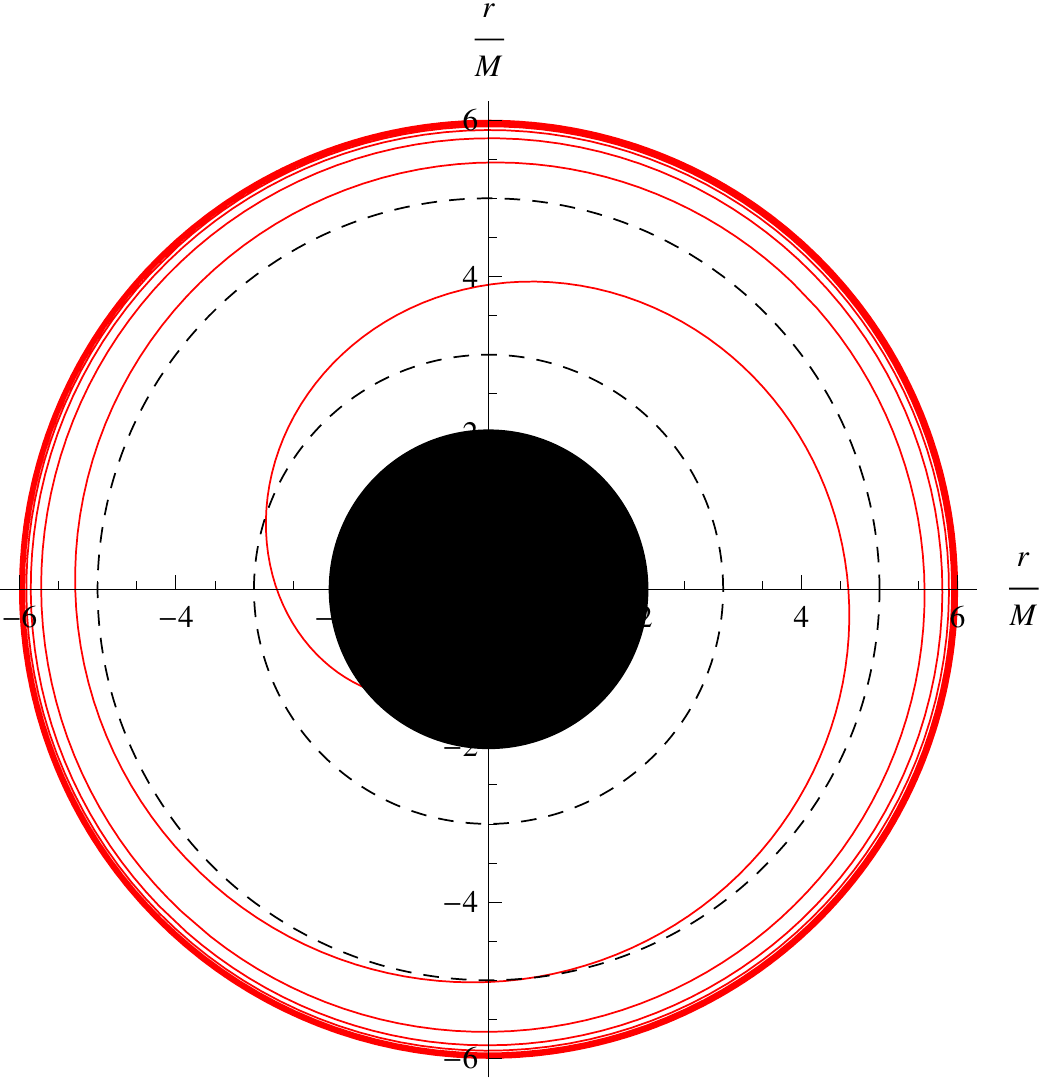}}
\vs{1}

{\footnotesize{Fig.\ 3: The inspiralling ballistic orbit towards the Schwarzschild horizon for $e = 0.0015$. }}
\ec

\section{Gravitational waves}

When a test mass $\mu$ falls towards the horizon on a ballistic orbit, it emits gravitational waves. In this section 
we compute the gravitational wave signal and establish that the total energy radiated is only a small fraction of the 
energy of motion of the test body; therefore the motion on such a geodesic should be a reliable approximation to the 
exact orbit of an infalling compact object. It is then of interest to compare these calculations in the limit of small 
$e$ with known results for the infall of a compact body from the ISCO. 

Our computation of the gravitational wave signal for EMR systems is based on the Zerilli-Moncrief and 
Regge-Wheeler equations \ct{RW,zerilli,moncrief} for the angular modes of the linearized gravitational-wave forms 
$(\psi^{lm}_{ZM}, \psi^{lm}_{RW})$:
\be
\label{n2.22}
\big(\partial_{r^*}^2-\partial_t^2-\bar{V}^l_{ZM/RW}(r^*)\big)\psi_{ZM/RW}^{lm} = \bar{S}^{lm}_{ZM/RW},
\ee 
where $r^*$ represents the standard radial `tortoise' co-ordinate
\[
r^* = r + 2M \ln \lh \frac{r}{2M} -1 \rh,
\] 
whilst in terms of the original $r$-co-ordinate
\bea 
\label{n2.23} 
\nonumber
\bar{V}_{ZM}^{l} &=& \left(\schwa\right)\frac{1}{r^2\Lambda^2(r)}\lh2\lambda^2(\lambda+1+\frac{3M}{r})
 + \frac{18M^2}{r^2}(\lambda+\frac{M}{r})\rh,\\ \nonumber
\bar{V}_{RW}^{l} &=& \left(\schwa\right)\frac{1}{r^2}\lh l(l+1)-\frac{6M}{r}\rh,\\ \nonumber
\lambda &=& \frac{(l+2)(l-1)}{2}\qquad\quad\Lambda(r) = \lambda+\frac{3M}{r} .
\eea
The expression for $\bar{S}^{lm}$ on the right-hand side of eq.\ (\ref{n2.22}) represents the angular modes of the 
sources for the gravitational waves generated by the motion of the compact object in the black-hole background. 
Their explicit form is discussed in the appendix. The gravitational-wave amplitudes observed at large distance from 
the source can be reconstructed from the solutions for $\psi^{lm}_{ZM/RW}$ through the complex combination
\be
\label{n2.28}
h_+(r,t,\theta,\phi) - ih_\times(r,t,\theta,\phi) = \frac{1}{r}\sum_{l,m}\left( \psi^{lm}_{ZM}(r^*_{obs},t)-2
 i\int_{-\infty}^t\psi^{lm}_{RW}(r^*_{obs},t')dt'\right){}_{-2}Y^{lm}(\theta,\phi),
\ee
with $(h_+, h_{\times})$ the two independent circular polarization modes of the waves in the $TT$-gauge, and 
$_{-2}Y^{lm}(\theta,\phi)$ the spin-weighted spherical harmonics. The rates at which energy and angular 
momentum are carried away by the gravitational waves are
\be
\ba{lll}
\label{n2.29}
\dsp{ \frac{dE}{dt} }& =& \dsp{ 
\frac{1}{64\pi}\sum_{l,m}\frac{(l+2)!}{(l-2)!}\left(|\dot{\psi}^{lm}_{ZM}(r^*_{obs},t)|^2 +  4|\psi^{lm}_{RW}(r^*_{obs},t)|^2\right), }\\
 & & \\
\dsp{ \frac{dL}{dt} }& = & \dsp{ 2\, \mbox{Re} \left[ \frac{i}{128\pi}\sum_{l,m}m\frac{(l+2)!}{(l-2)!} 
  \left(\dot{\psi}^{lm}_{ZM}(r^*_{obs},t)\psi^{*lm}_{ZM}(r^*_{obs},t) \rd \rd }\\
 & & \\
 & & \dsp{ \ld \ld \hs{3} +\, 4\psi^{*lm}_{RW}(r^*_{obs},t)\int_{-\infty}^t \psi^{*lm}_{RW}(r^*_{obs},t')dt' \right) \right]. }
\ea
\ee
We have computed numerical solutions for the angular modes $\psi^{lm}_{ZM/RW}$ and the resulting metric perturbations 
$h_{+,\times}$, using a C$^{++}$-implemented version of the Lousto-Price algorithm \cite{MartelPoisson,lousto-price}. 
The algorithm is based on dividing the $(t,r^*)$ surfaces in Schwarzschild space-time into cells of equal area, which also 
represent cells of equal physical area. The differential equations (\ref{n2.22}) are discretized on this grid, and solutions are 
computed from appropriate initial conditions, such that further refinement of the grid does not change the result up to the 
desired accuracy. 

To obtain a reliable wave signal for the infall on a ballistic orbit starting close to the ISCO from a quasi-periodic orbit,
we consider a Schwarzschild space-time in which the inspiral of the test mass has produced a continuous set of outgoing 
gravitational waves. Realistic initial conditions for the angular modes $\psi^{lm}_{ZM/RW}$ are created by letting the test 
mass run on the ISCO for a sufficiently long period that any initial transient waves have passed the point where the observer 
is located and the wave forms are measured. We take this point to be in the equatorial plane at $r = 500 M$. 

Next we set the compact mass on a nearby ballistic orbit, and compute the resulting angular wave modes and metric 
perturbations. The instantaneous shift in the orbit, even if small, also produces a transient signal. However, for small 
$e$ the ballistic orbit has an almost-periodic initial stage during which it runs close to the ISCO for a long time. During  
this time the original periodic signal is recovered to a very good approximation. 

In fact, initial transients are common in standard numerical techniques, arising for example from the well-known Gibbs 
phenomenon \ct{HopperEvans, Nagar-Tartaglia}. We have checked that changing the initial conditions in various ways, 
though affecting the transient signal, does not change the almost-periodic signal during the initial stage of the ballistic orbit, 
nor the development of the signal afterwards. Thus it is confirmed that, as in other numerical works \ct{Nagar, Bernuzzi},  
spurious transients appear only in the initial phase of the signal and are followed by numerically stable physical wave 
forms.  

Fig.\ 4 shows the results for the most relevant ZM- and RW-modes, as calculated for a system with mass ratio 
$\nu = \mu/M = 10^{-7}$ on a ballistic orbit with $e = 1.5 \times 10^{3}$, as observed from the equatorial plane. 
For convenience of representation the amplitudes have been rescaled by the inverse of the mass ratio. 

\bc
\scalebox{0.47}{
                  \includegraphics{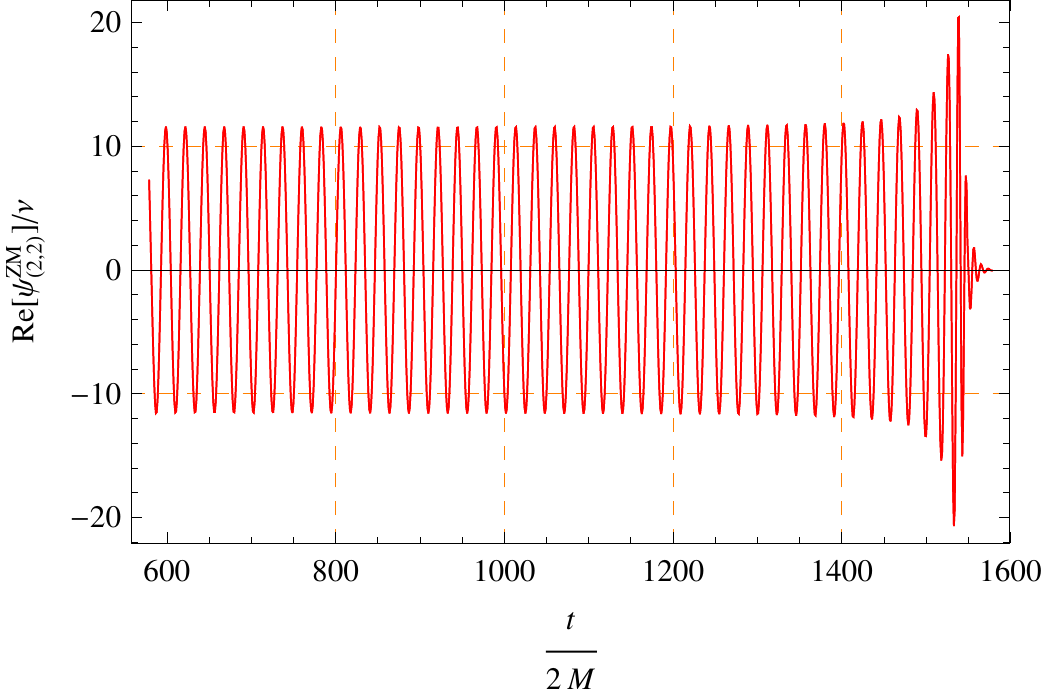} \hs{1}
                  \includegraphics{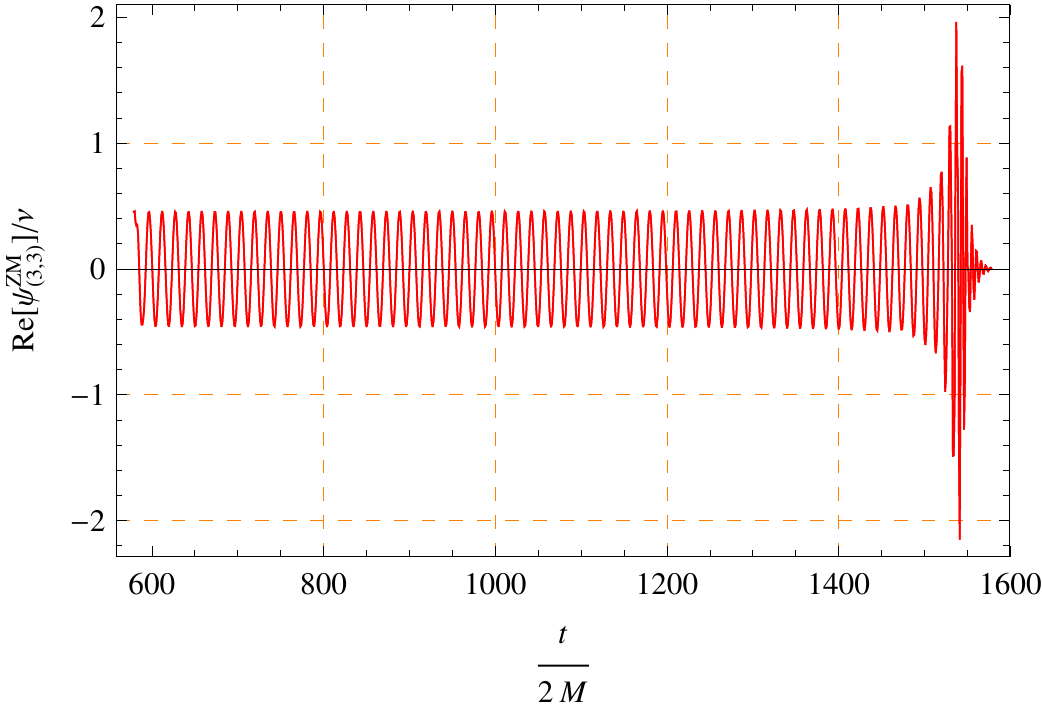} \hs{1}
                  \includegraphics{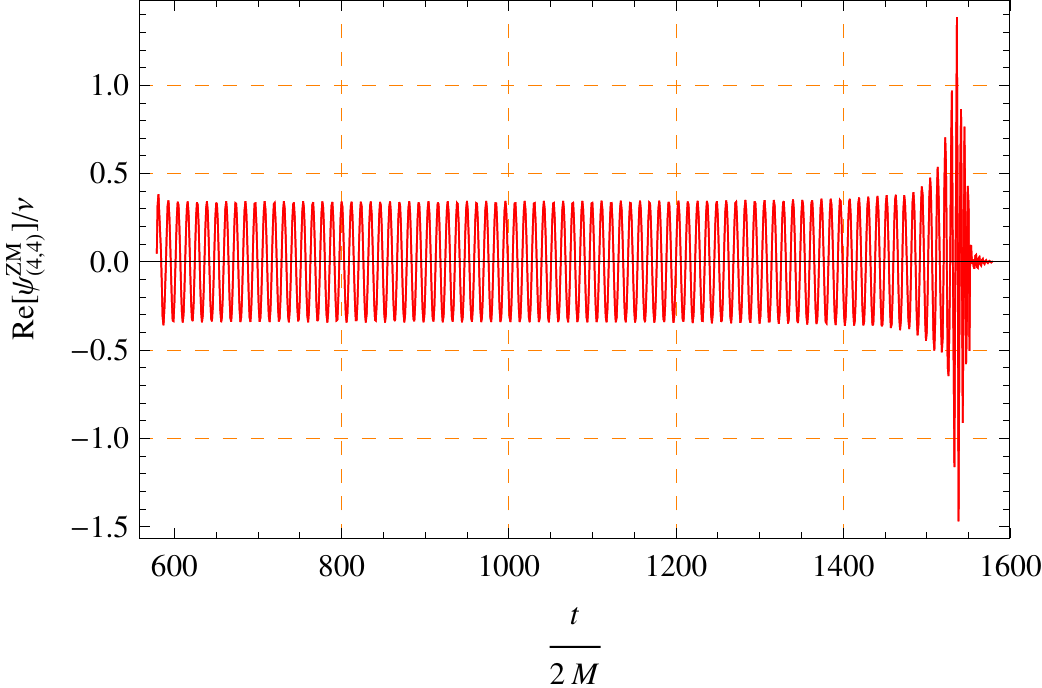} }        
           
\scalebox{0.47}{                
                  \includegraphics{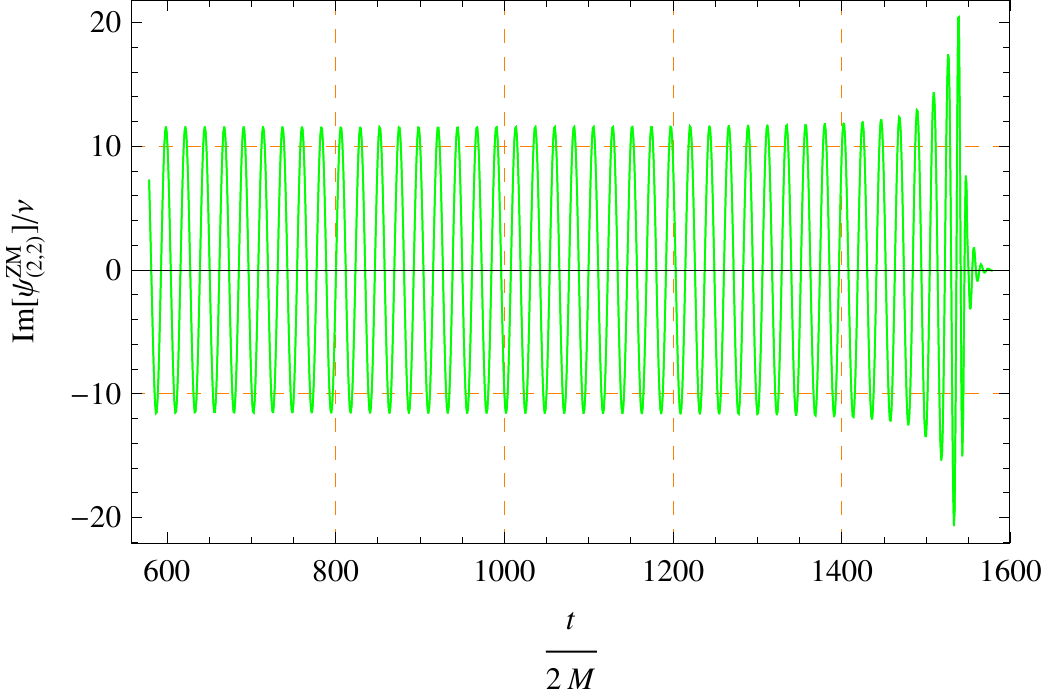} \hs{1}
                  \includegraphics{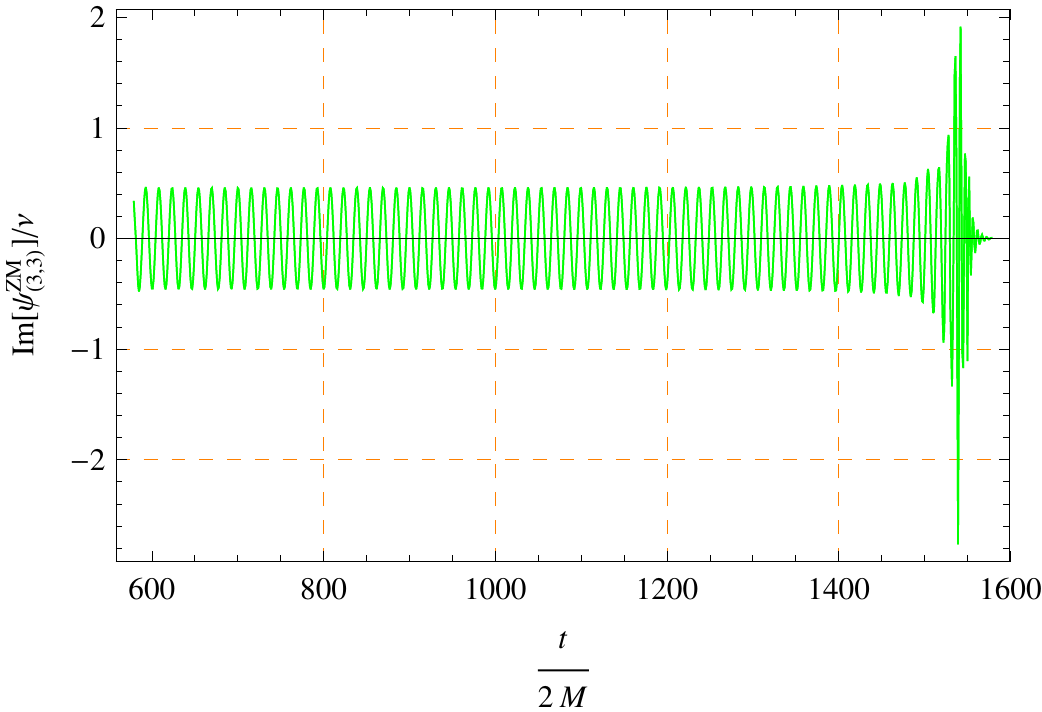} \hs{1}
                  \includegraphics{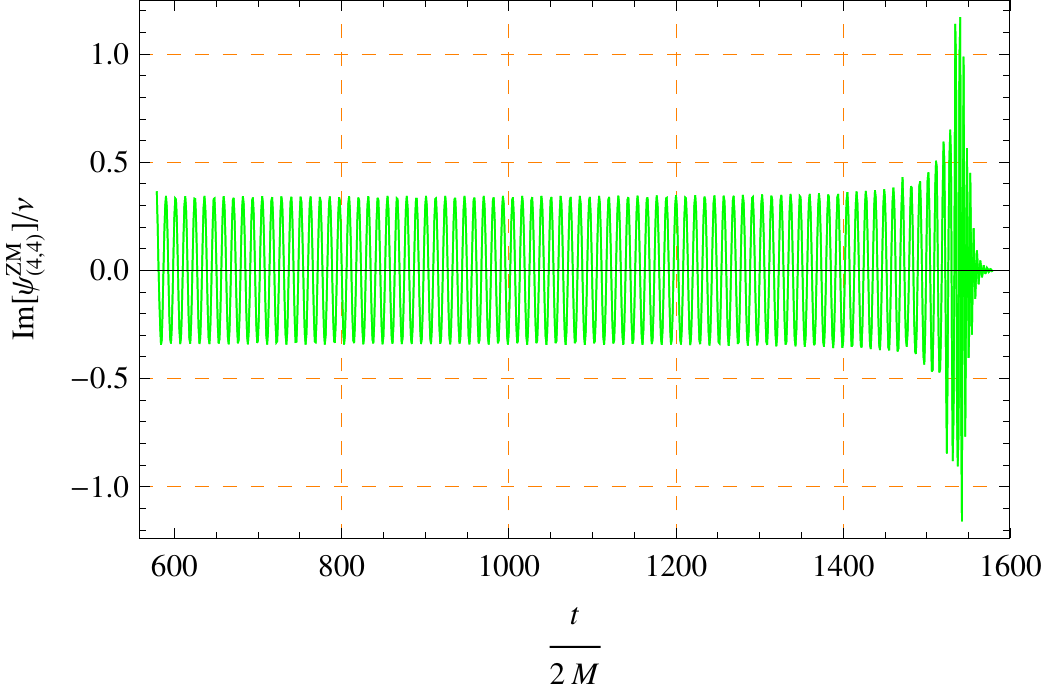} }
           
\scalebox{0.47}{
                  \includegraphics{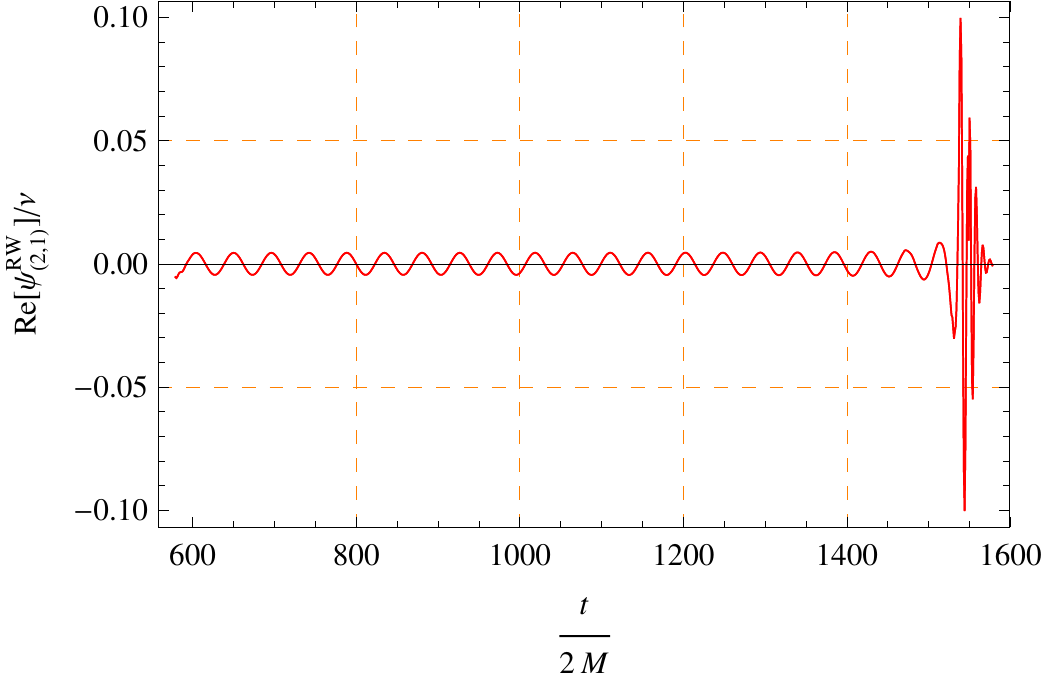} \hs{1}
                  \includegraphics{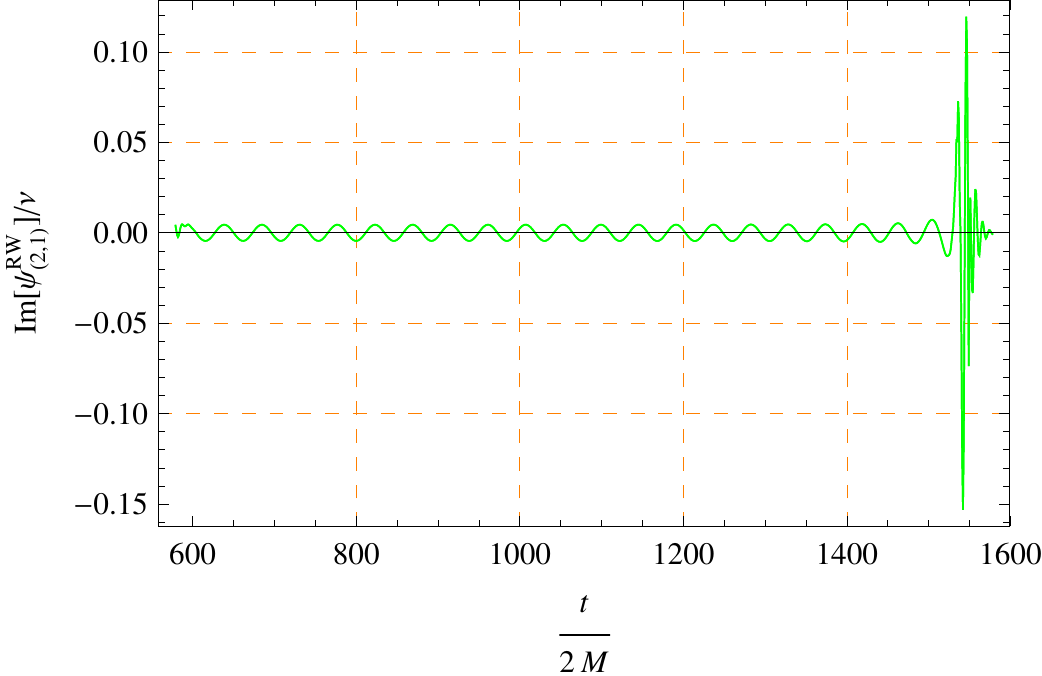} \hs{1}
                  \includegraphics{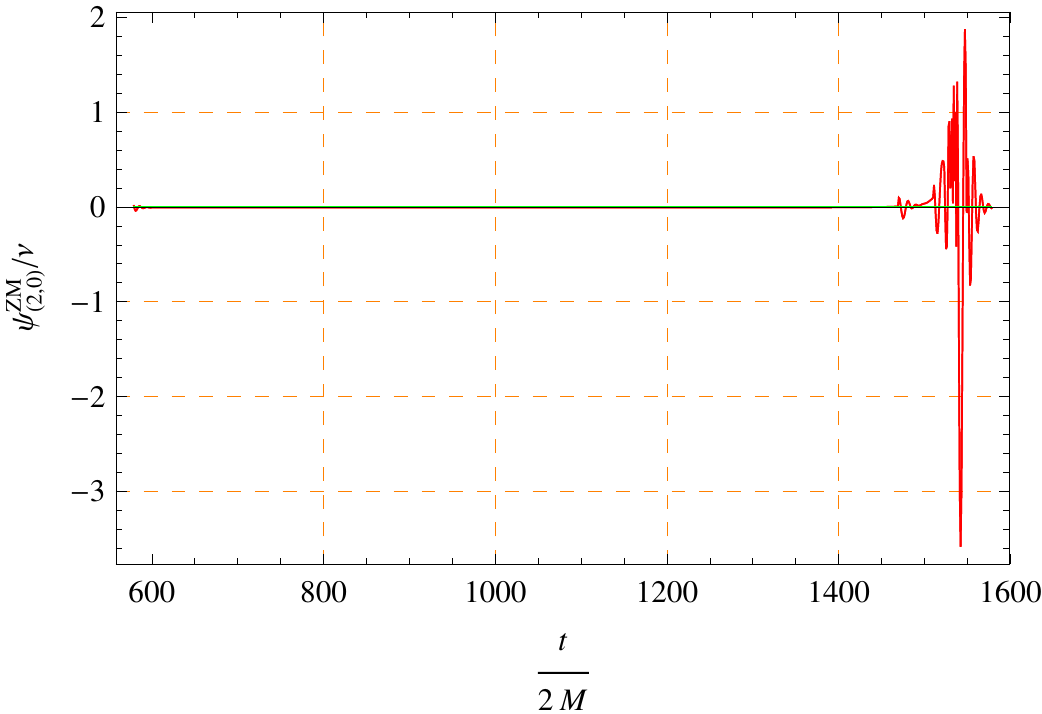} }     
 \vs{1}
 
     {\footnotesize{Fig.\ 4:  Top row: Re parts (red) of the amplitudes $\psi^{ll}_{ZM}$ for $l = 2,3,4$, magnified by $1/\nu$. \\
     \hs{-8.8} Middle row: Im parts (green) of the same amplitudes. \\
     \hs{2.5} Bottom row: Re and Im parts of the amplitude $\psi^{21}_{RW}$, and real amplitude $\psi^{20}_{ZM}$. }} 
 \ec

\nit
Both the real and imaginary parts are shown, from the time after the spurious transient signals have disappeared. 
The mode with $m = l = 2$ dominates all others; the modes with $m = l \geq 3$ are subdominant, representing the 
most important corrections. The modes with $m < l$ are  negligeable, even for $l = 2$.

Inserting the results for the $\psi^{lm}_{ZM/RW}$ into expression (\ref{n2.28}) for the gravitational wave amplitude 
produces the wave forms shown in figs.\ 5 and 6, as seen from two orthogonal directions in the equatorial plane.
The figures show that for about 20 revolutions the signal is almost periodic, ending in a short burst of radiation. 
The transition from almost-periodic to burst happens in the region where the cross-over from nearly circular 
to strongly radial motion occurs. Once the mass $\mu$ crosses the light ring at $r = 3M$ the gravitational 
waves apparently red-shift and quickly fade.

The transition in the gravitational wave signal of the ballistic orbit from almost periodic to burst can be identified 
also in the emitted energy and angular momentum, as is clear from figs.\,7. The power peaks at time $t = 1536.5 M$, 
shortly before the compact mass crosses the light ring. The total energy emitted during the infall on a ballistic orbit is a 
fraction $0.33 \times 10^{-4}$ of the original energy of motion of the infalling mass. The back reaction from such a small 
energy loss would only require a very tiny correction to the geodesic motion, as anticipated.
\vs{1}

\bc
\scalebox{0.6}{                 
          \includegraphics{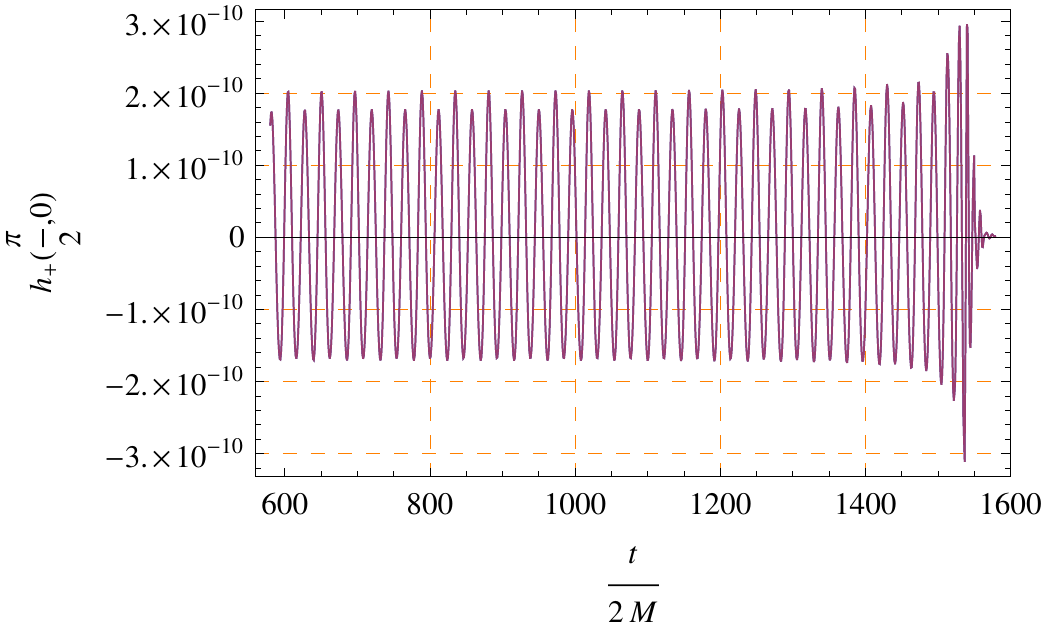} \hs{1}
          \includegraphics{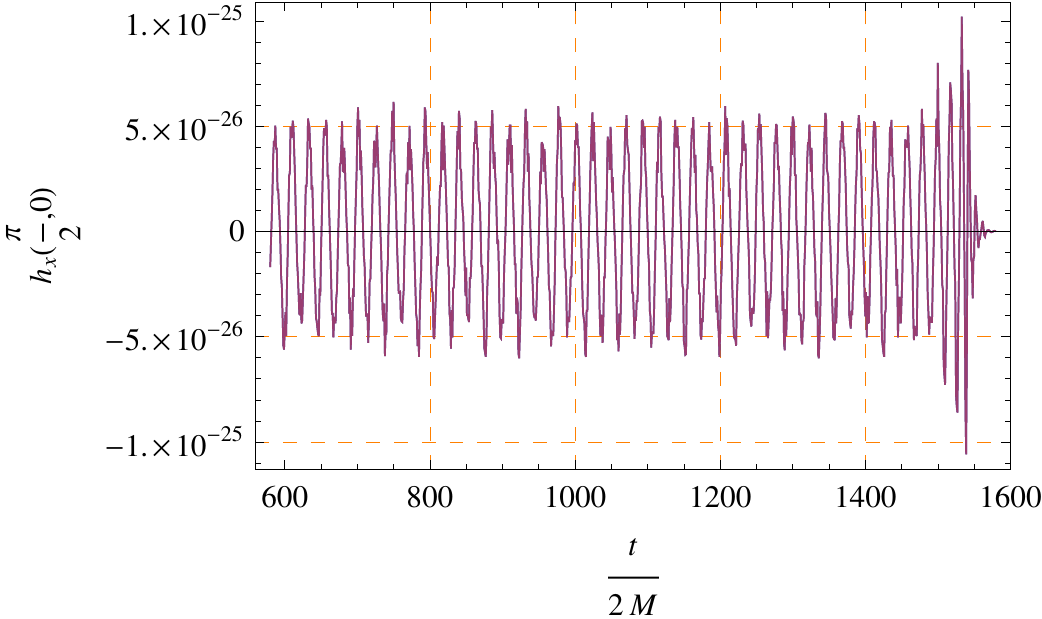} }
\vs{1}

 {\footnotesize{Fig.\ 5: $h_+(\frac{\pi}{2},\varphi = 0), h_\times(\frac{\pi}{2},\varphi = 0)$ polarizations of gravitational 
  waves radiated by a system with mass ratio $\nu = 10^{-7}$ on a ballistic orbit for $e = 0.0015$.}}
\ec

\bc
\scalebox{0.6}{   
         \includegraphics{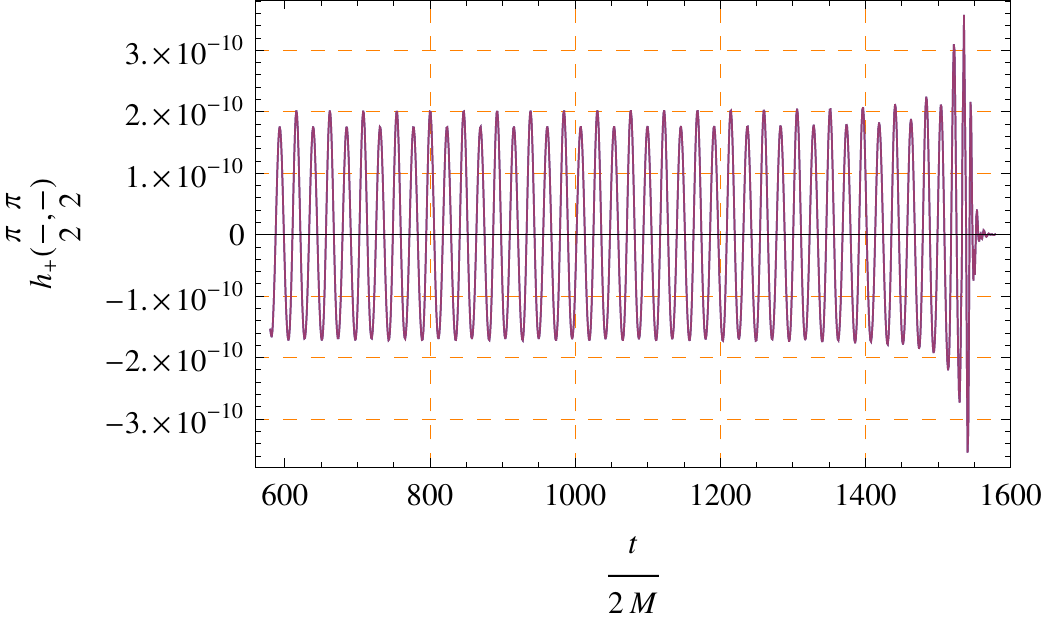} \hs{1}
         \includegraphics{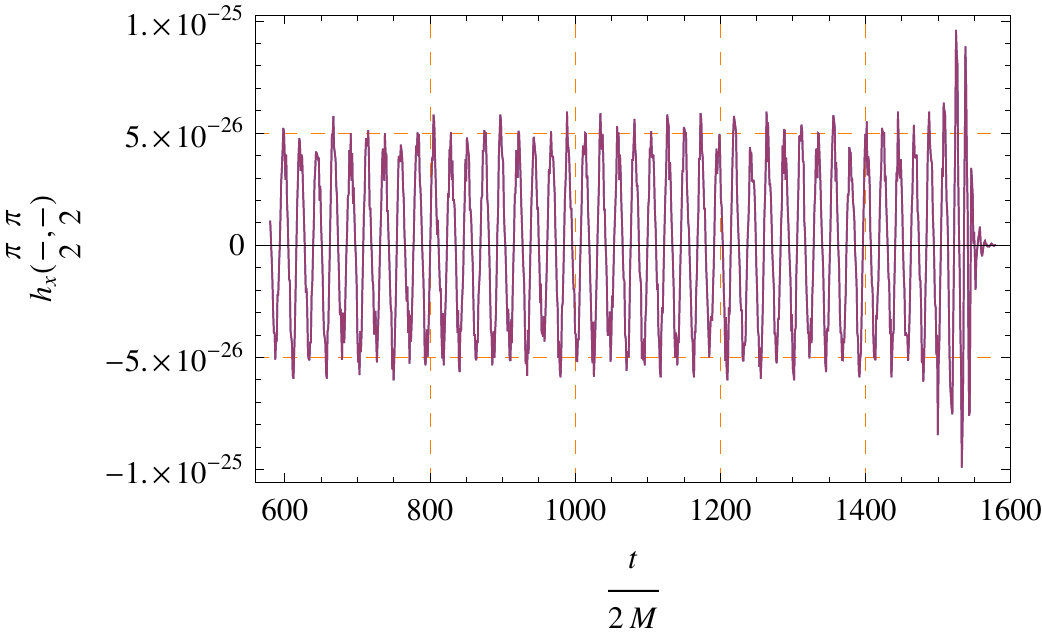} }
\vs{1} 

 {\footnotesize{Fig.\ 6: $h_+(\frac{\pi}{2},\varphi = \pi/2), h_\times(\frac{\pi}{2},\varphi = \pi/2)$ polarizations of 
 gravitational waves  radiated by a system with mass ratio $\nu = 10^{-7}$ on a ballistic orbit for $e = 0.0015$.}}
\ec
\vs{2}

\bc
\scalebox{0.7}{
         \includegraphics{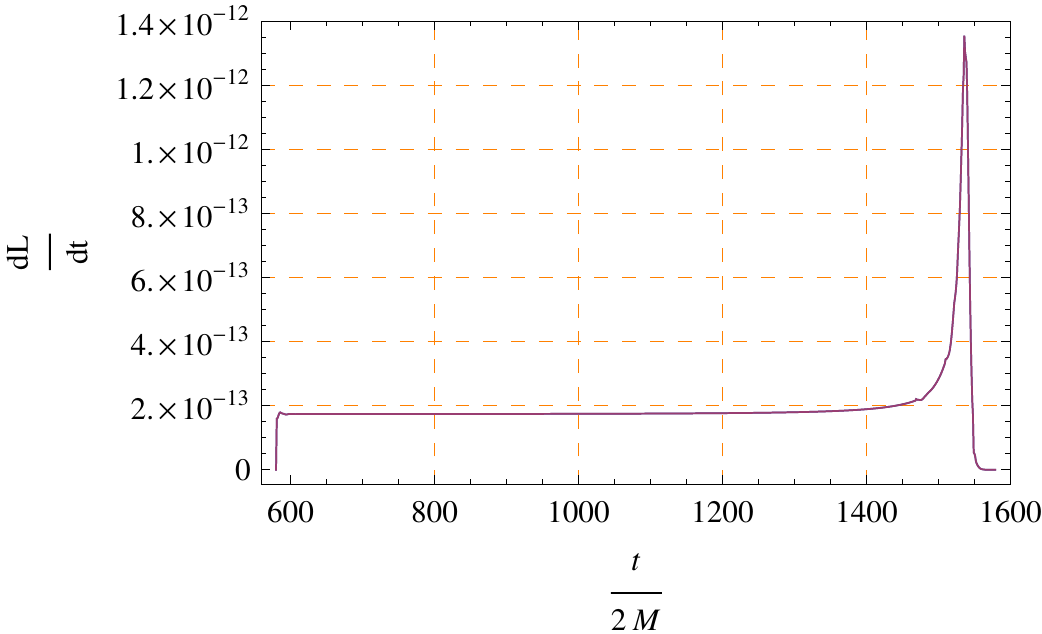} \hs{1}
         \includegraphics{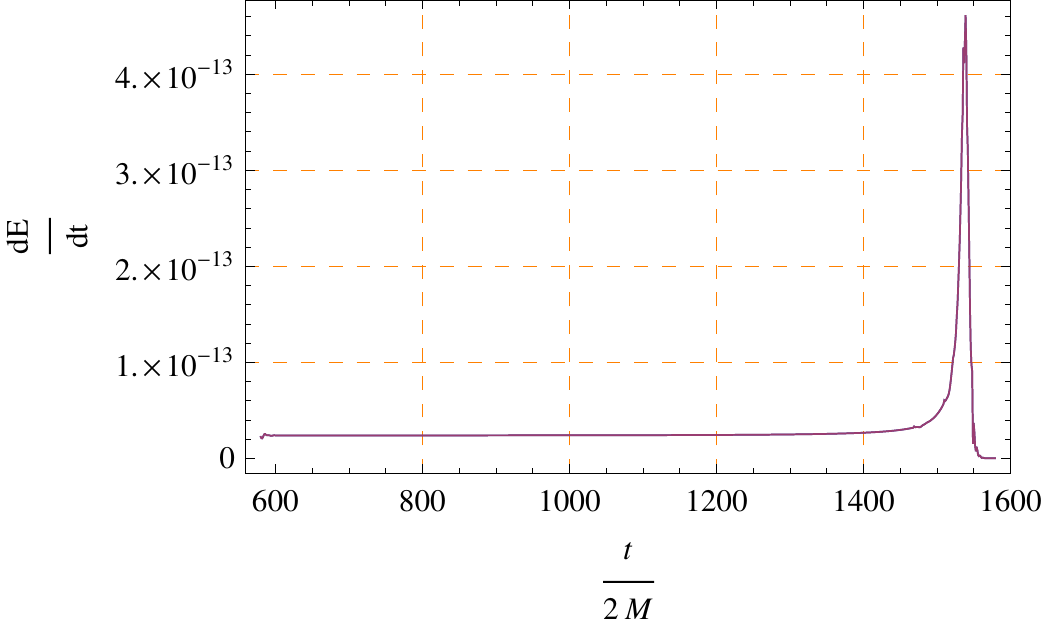} }
\vs{1}

{\footnotesize{Fig.\ 7: Angular momentum loss (left) and energy loss (right) on a ballistic orbit with $e = 0.0015$.}}
\ec

\section{Discussion}

The ballistic orbits (\ref{2.7}), (\ref{2.8}) are  exact solutions of the geodesic equation for Schwarzschild space-time. 
They describe the motion of a test mass rising up from the horizon to a maximum radial distance $R < 6M$, before
returning to the horizon; for $R$ close to $6M$ this includes a large number of turns around the black hole. 
As they can come arbitrarily close to the ISCO and are degenerate in energy and angular momentum with stable 
circular orbits, in the limit of small $e$ they can be used to describe test masses infinitesimally boosted from the 
ISCO to the infalling part of a ballistic orbit. For such infalling test masses we have computed the gravitational 
waves emitted during the infall. 

\bc
\scalebox{0.65}{ \includegraphics{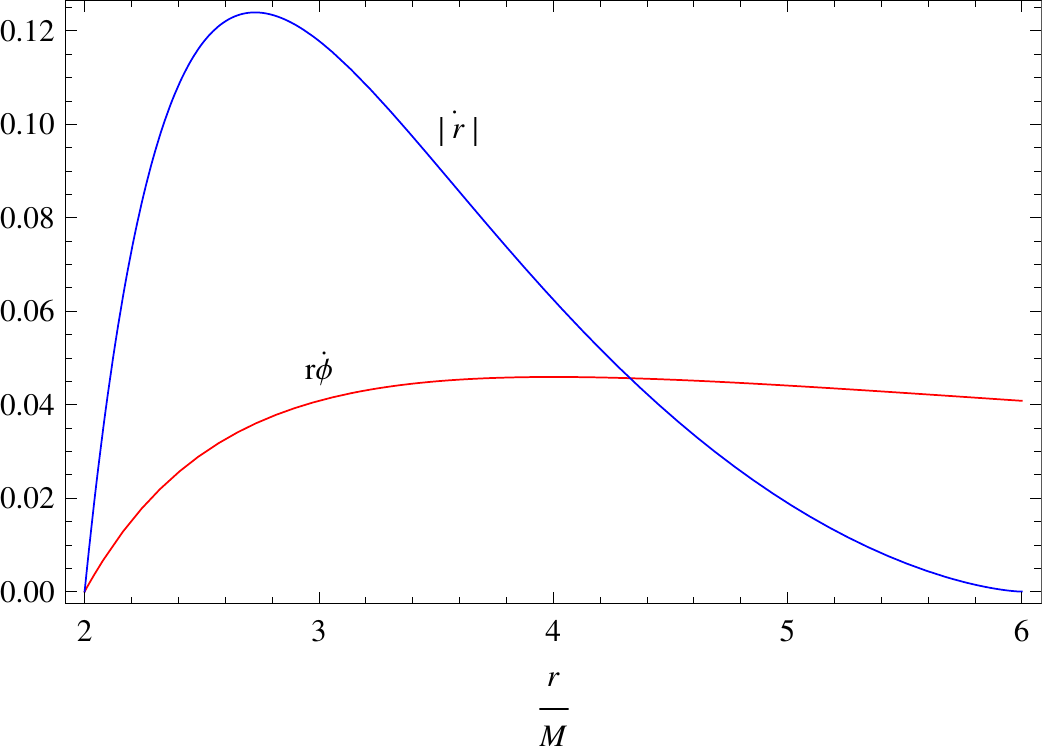} }
\vs{1}

{\footnotesize{Fig.\ 8: Evolution of circular and radial velocities as a function of $r$ for the orbit with $\delta=0.001$;
the cross-over occurs at $r_* = 4.328 M$.}}
\ec

The merger phase of ballistic EMRs, characterized by an energy and an angular momentum only slightly higher 
than those on the ISCO, is divided into two phases. First the point mass $\mu$ follows an almost-circular orbit, for 
which $\dot{r}\ll r\dot{\varphi}$.  Near $r = 4.3 M$ the radial and circular velocities become equal, and after  
this cross-over the radial motion dominates; see fig.\ 8. This subdivision in the orbit was already recognized in 
ref.\ \ct{Buonanno}, when studying more general cases of black-hole coalescences. 

Our calculations show, that the number of almost circular revolutions depends strongly on the radial distance 
of the apastron, the point of closest approach to the ISCO. It is convenient to parametrize this radial co-ordinate 
by $R = 6M (1 - \del)$; equivalently 

\be
\delta = \frac{2}{3} e.
\label{delta}
\ee

\nit
In contrast the radial distance $r_*/M$ where the cross-over occurs, and the number of turns during the 
second phase of radially dominated motion (the change in orbital phase in units of $2\pi$) are virtually 
independent of $\del$. These conclusions are illustrated by the numbers in table 1, calculated from our 
analytic formulae. 
\vs{1}

\bc
\begin{tabular}{|r|r|c|c|}\hline
$\del \times 10^3$ &  $r_*/M$ & $n_1$ & $n_2$  \\ \hline
1.3 & 4.328 & 18.71 & 0.49 \\
3.2 & 4.328  & 11.59 & 0.50 \\
7.8 & 4.328 & 7.09 & 0.50 \\
18.3 & 4.327 & 4.29 & 0.50 \\
41.2 & 4.323 & 2.53 & 0.50  \\ \hline
\end{tabular}
\vs{1}

{\footnotesize Table 1: Circular vs.\ radial phase on ballistic orbits; $r_*$ is the radial co-ordinate at which the
cross-over occurs; $n_1$ is the number of turns in the orbit before reaching the cross-over; $n_2$ is the 
number of turns between cross-over and horizon.}
\ec

\nit
The first phase is characterized by almost-periodic gravitational radiation, very similar to the waves emitted
by a compact mass in orbit on the ISCO. The second phase instead is much shorter and characterized by 
a burst of gravitational radiation in which the amplitude increases by a factor of two before the  mass 
disappears effectively behind the light ring.

One can calculate from expressions like (\ref{n3.5}) the time elapsed during plunge, from the apastron 
to the horizon $r = 2M$, and this time can vary in a range of values $\frac{t}{2M} \in [60,1000]$ for 
$\delta \in [10^{-3},10^{-1}]$. Obviously this range of values is determined largely by the duration of 
the nearly-circular phase of the orbit. It is also evident from the gravitational-wave signal; a comparison 
of the dominant component of the Regge-Wheeler modes for various values of $\delta$ as in fig.\ 9 shows 
quite similar final bursts of radiation, whilst the duration of the almost-periodic part of the signal varies 
considerably \cite{Nagar,Scheel,Damour}.

\bc
\scalebox{0.47}{
        \includegraphics{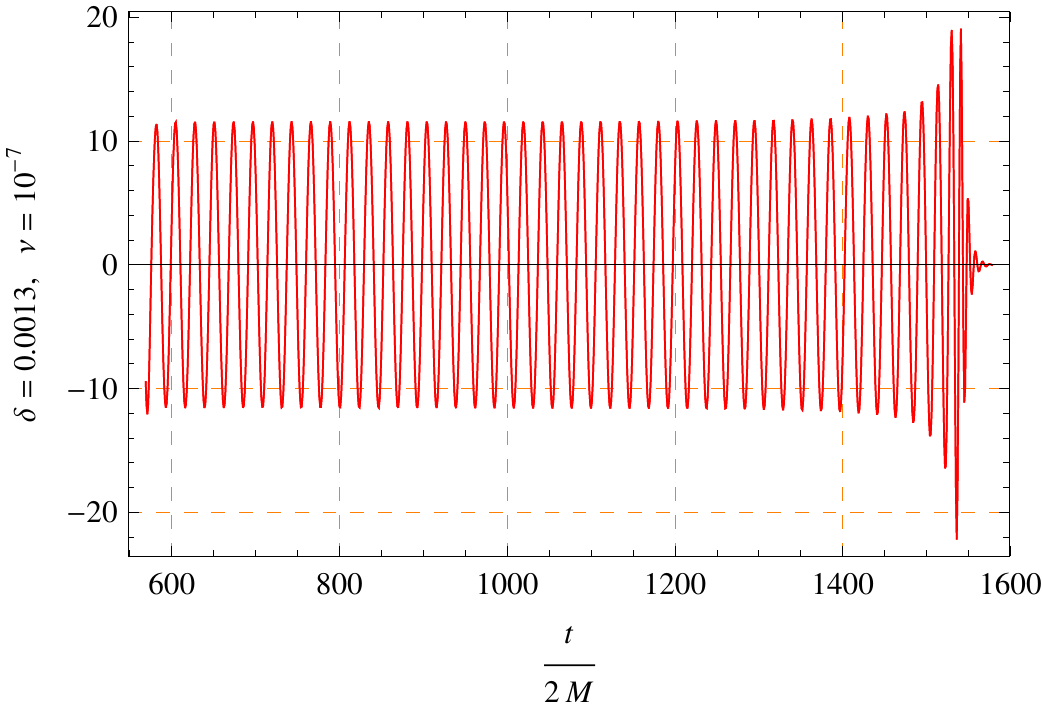} \hs{1}
        \includegraphics{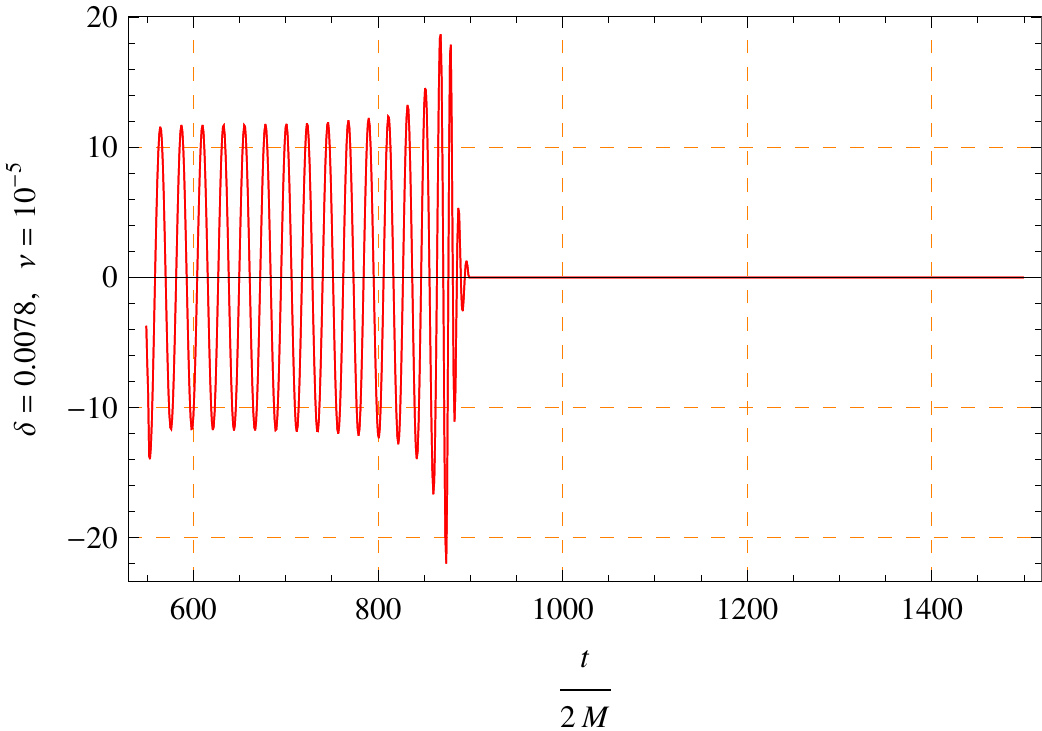} \hs{1}
        \includegraphics{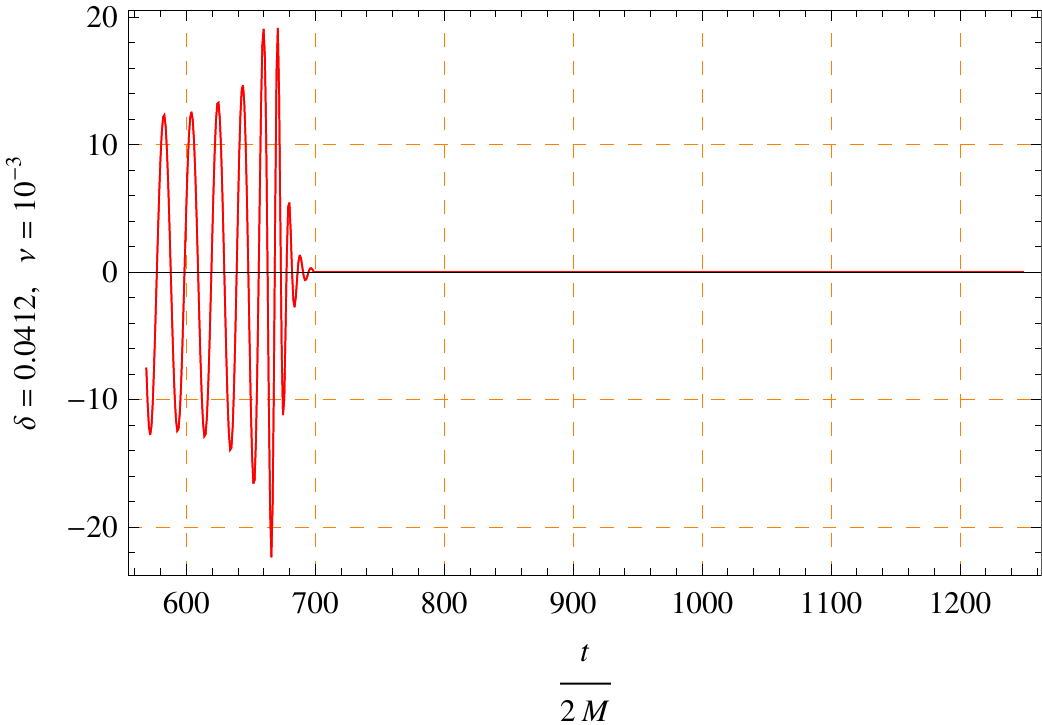} }
\vs{1}  
             
      {\footnotesize{ Fig.\ 9: $\psi^{(2,2)}_{ZM}/\nu$ amplitudes for $\delta = (0.0013, 0.0078, 0.0412)$, only real 
         parts shown. The values of $\delta$ are chosen to match mass ratios $\nu = (10^{-7}, 10^{-5}, 10^{-3})$. 
	 The smaller $(\nu, \del)$, the longer the near-circular \\ phase after $\frac{t}{2M}=550$.}}
\ec 
In specific parameter ranges the ballistic orbits can be used as a first approximation to the infall phase of an 
EMR binary. Comparing the wave forms for ballistic orbits with those computed for direct infall from the 
ISCO\footnote{The ISCO is commonly regarded to be the region where the transition from adiabatic inspiral 
and plunge takes place \ct{Buonanno,Bernuzzi,Damour}.},  based on Post-Newtonian and Effective One-Body 
approximations \ct{Nagar}-\ct{Damour}, it is easy to see that they share the same qualitative behaviour. 

For a more precise quantitative comparison we have to tune $\delta$ carefully. In fact in \cite{Buonanno,Damour} 
it has been shown that the number of revolutions during plunge from the ISCO for coalescences is roughly 
$\sim (4\nu)^{-1/5}$, where $\nu = \frac{\mu}{M}$ is the mass ratio. The mechanism behind this dependence is 
the fractional loss of energy by emission of gravitational radiation. As the ballistic orbits are exact geodesics, 
they do not account for such energy loss, but we can tune $\delta$ to make up for this and obtain a realistic 
approximation. For example, $\nu = 10^{-7}$ corresponds to $19$ revolutions and $\delta = 0.0013$. In 
fig.\ 9 the mode amplitude for $l = m = 2$ is shown for values of $\del$ matching with 
$\nu = (10^{-7}, 10^{-5}, 10^{-3})$, respectively.

As a check on these results, we can compare with another result of \ct{Buonanno}, elaborated in \ct{Nagar}, 
stating that the onset of universal behaviour starts inside the ISCO at co-ordinate distance 
\be
\frac{r}{M} = 6 - \ag \nu^{2/5},
\label{nu}
\ee
where $\ag$ is a constant. As the ballistic orbits are geodesic and universal, we should expect the values of 
$\del$ to scale in the same way: $\del = \gam \nu^{2/5}$. We find that this relation holds reasonably well with 
$\gam = 0.8$ for small values $\nu \leq 10^{-5}$, beyond which deviations arise of the order of 10\% or more. 
Table 2 summarizes the numerical results for this comparison.
\vs{1}

\bc
\begin{tabular}{|r|r|c|}\hline
$\del \times 10^3$ & $\nu$ &  $\gam = \nu^{-2/5} \del$ \\ \hline
1.3 & $10^{-7}$ & 0.82 \\
3.2 & $10^{-6}$ & 0.80 \\
7.8 & $10^{-5}$ & 0.78 \\
18.3 & $10^{-4}$ & 0.73 \\
41.2 & $10^{-3}$ & 0.67 \\ \hline
\end{tabular}
\vs{1}

{\footnotesize Table 2: Test of the universality of the dependence of $\del$ on the mass ratio $\nu$.}
\ec

\nit
Deviations for larger mass ratios are to be expected for at least two reasons: the center-of-mass 
motion and back reaction of the compact object on the Schwarzschild background geometry have 
been neglected in our treatment; and the kinetic energy of the compact object is overestimated according 
to eq.\ (\ref{2.10}) by a factor $\Del \ve/\ve = 3\del/8$.

For ballistic orbits in EMR binaries -- similar to what is believed to happen for more general cases -- the total energy 
emitted during infall is only a tiny fraction of the initial energy. Therefore in practice we expect the deviation from 
geodesic motion to be small. Indeed, it has been argued in the literature before \ct{Buonanno} that in the particle 
limit $\nu\rightarrow 0$ the motion is driven mainly by the central black hole, leading to quasi-geodesic motion.  
Moreover, the energy emission is far too weak to exceed the potential barrier created 
by the Zerilli potential (\ref{n2.23}), which shows a maximum in the region close to the light-ring $r\simeq 3M$. 
Thus the final burst of gravitational waves reaching a distant observer occurs before the compact mass crosses this 
barrier. No substantial amount of radiation coming from the inner region $r<3M$ can be transmitted to the outer region 
\ct{Bernuzzi,berti_et-al,Davis, bnz}.

To summarize, we have explored the nature of ballistic orbits in Schwarzschild space-time, and established their 
degeneracy in the $(\ve, \ell)$-plane with circular orbits. We have computed gravitational wave signals for compact 
objects falling towards the horizon on such an orbit. We have also argued that ballistic orbits with apastron close to 
$6M$ can provide good first approximations to infall from bound orbits, in particular from the ISCO. Finally, we have 
confirmed earlier calculations that the total amount of energy converted into gravitational waves during the plunge 
phase is small, and that a quasi-geodetic description of this process for EMR binaries seems adequate. 

The results reported here can be improved further by computing perturbative corrections to the orbits by the method of 
geodesic deviations, as was done for the quasi-periodic motion during the inspiral phase in refs.\ \ct{GKJWvH,GKJWvH2}.
\vs{3}

\nit
{\bf Acknowledgements}\\
The authors acknowledge useful correspondence with Maarten van de Meent of the University of Southampton.  
The research reported in this paper is supported by the programme on Gravitational Physics of the Foundation 
for Fundamental Research of Matter (FOM).

\np
\appendix
\section{Source terms in the wave equation}

The expressions for the source terms in the Zerilli-Moncrief and Regge-Wheeler equations for an infalling mass 
on a ballistic orbit can be evaluated using the decomposition into tensor-spherical harmonics as described in 
detail in refs.\ \ct{Martel} and \ct{koekoek}. In the point-particle limit the source terms are concentrated on the 
world-line of the particle, producing delta-functions and their derivatives. In Schwarzschild space-time the 
angular modes then take the form
\be
\bar{S}^{lm}_{ZM/RW} =  \big(\schwa\big)\big(F^{lm}_{ZM/RW}\partial_r\delta(r-r_p)+G^{lm}_{ZM/RW}\delta(r-r_p)\big) .
\label{a.2}
\ee
The expressions for the coefficient functions $F^{lm}$ and $G^{lm}$ are provided below for the $ZM$ and $RW$ 
cases separately. For computational convenience the results are given in terms of the rescaled eccentricity
$\del = {2e/3}$, as defined in eq.\ (\ref{delta}).
\vs{1}

\nit
{\em Zerilli-Moncrief modes} \\
The even modes of the gravitational wave potentials are calculated from the Zerilli-Moncrief equation; the
corresponding source terms are
\be
\ba{l}
 F^{lm}_{ZM}(r,\delta,\varphi) = \\
 \\
\dsp{ \hs{3} \frac{-24\pi\mu Y^{*lm}(\frac{\pi}{2}, \varphi)}{(\lambda + 1)\Lambda}
  \left[ \frac{(2M-r)^2\sqrt{2-3\delta+\delta^2}(48M^2(\delta-1)^2
  +r^2(4-8\delta+3\delta^2))}{r^4(9\delta^3-36\delta^2+44\delta-16)} \right], }
\ea  
\ee
and
\be
G^{lm}_{ZM}(r,\delta,\varphi) = g_1Y^{*lm} \lh \frac{\pi}{2}, \varphi \rh + g_2Z_\varphi^{*lm} \lh \frac{\pi}{2}, \varphi \rh 
 + g_3U_\varphi^{*lm} \lh \frac{\pi}{2}, \varphi \rh + g_4V_{\varphi\varphi}^{*lm}  \lh \frac{\pi}{2}, \varphi \rh, 
\ee
with $g_i = \frac{N_i}{D_i}, i=1,2,3$ defined by

\nit
\[
\ba{lll}
 N_1 &  = & \dsp{ -8\pi\mu\big(\schwa\big)(3\delta-4)\sqrt{2-3\delta+\delta^2}\Bigg[3M(5M-3r)r^2(4-3\delta)^2(3\delta-2) } \\
 & & \\
 & & \dsp{ +\, \quad(4M-r)r^3(4-3\delta)^2(3\delta-2)\lambda+r^4(4-3\delta)^2(3\delta-2)\lambda^2 } \\
 & & \\
 & & \dsp{+\, \quad\Lambda\Bigg(8640M^4(\delta-1)^3-6912M^3r(\delta-1)^3+2Mr^3(112-336\delta+306\delta^2-81\delta^3) } \\ 
 & & \\
 & & \dsp{ \quad+\, 108M^2r^2(15\delta^3-47\delta^2+48\delta-16)+r^4(54\delta^3-189\delta^2+204\delta-68)\Bigg) } \\
 & & \\
 & & \dsp{ \quad -\, r(r+6M(\delta-1))(12M(1-\delta)+r(3\delta-2))^2\Lambda^2\Bigg], } \\ 
 & & \\
D_1 & = & \dsp{ 3(2M-r)r^4(4-3\delta)^2(\delta-2)(\delta-1)(3\delta-2)(1+\lambda)\Lambda^2, }
\ea
\]

\nit
\[
\ba{l}
N_2  \dsp{ = -128\sqrt{3}\pi\mu(r-2M)M\sqrt{(r+6M(\delta-1))(\delta-2)}(\delta-1)\lh12(1-\delta)M+r(3\delta-2)\rh, } \\ 
  \\
D_2  \dsp{ = l(l+1)r^3\sqrt{r^3}(3\delta-4)(4-8\delta+3\delta^2)\Lambda, } \\ 
  \\
N_3 \dsp{ = -1152M^2\pi\mu(r-2M)^2(\delta-1)^3, } \\ 
  \\
D_3 \dsp{ =r^6(1+\lambda)\Lambda(3\delta-4)(3\delta-2)\sqrt{2-3\delta+\delta^2}, } \\ 
  \\
N_4 \dsp{ = - 4608M^2\pi\mu(2M-r)(\delta-1)^3, } \\
  \\
D_4 \dsp{ = r^4(3\delta-4)(3\delta-2)\sqrt{2-3\delta+\delta^2}\, \frac{(l-2)!}{(l+2)!}. }
\ea
\]

\nit
{\em Regge-Wheeler modes}\\
The odd modes of the gravitational wave potentials are calculated from the Regge-Wheeler equation; 
the corresponding source terms are
\be
F^{lm}_{RW}(r,\delta,\varphi) = \frac{2304M^2\pi\mu(\delta-1)^2\sqrt{2-3\delta + \delta^2}(r-2M)^2(l-2)!}{r^5(9\delta^3-36\delta^2 +
 44\delta-16)(l+2)!}\, W_{\varphi\varphi}^{*lm}\lh \frac{\pi}{2}, \varphi \rh,
\ee
and 
\be
\label{2.27}
\ba{l}
G^{lm}_{RW}(r,\delta,\varphi) =  \dsp{ 
 -\, M\pi\mu \left[ \frac{9216M(r-3M)(r-2M)(\delta-1)^2(l-2)!}{r^6(9\delta^3-36\delta^2 + 
 44\delta-16)(l+2)!}\, W^{*lm}_{\varphi\varphi}\lh \frac{\pi}{2}, \varphi \rh \rd }\\
 \\
 \dsp{ \ld \hs{2} +\, \frac{64\sqrt{3}(r-2M)(\delta-1)^2(12M(1-\delta)+r(3\delta-2))\sqrt{\frac{r+6M}{r^3(\delta-2)(\delta-1)}})}{l(l+1) r^4 
 (3\delta-4)(3\delta-2)}X^{*lm}_\varphi\lh \frac{\pi}{2},\varphi \rh \right]. }
\ea
\ee

\np

\end{document}